\RequirePackage{amsmath}
\documentclass[runningheads]{llncs}

\usepackage{fancyvrb}
\usepackage[usenames]{color}
\usepackage{amssymb}
\usepackage{amsmath}
\usepackage{amsfonts}
\usepackage{amscd}
\usepackage{graphicx}
\usepackage{hyperref}

\usepackage{graphics}
\usepackage{latexsym}
\usepackage{epsf}

\DeclareMathOperator{\Fac}{Fac}

\newcommand{\seqnum}[1]{\href{https://oeis.org/#1}{\underline{#1}}}

\begin{document}
\title{Critical exponent of infinite balanced words via the Pell number system}
\titlerunning{Critical exponent of infinite balanced words}
%
\author{
Aseem R. Baranwal \orcidID{0000-0001-5318-6054}
\and
Jeffrey Shallit \orcidID{0000-0003-1197-3820}
}

\authorrunning{A. R. Baranwal and J. Shallit}
%
\institute{
    School of Computer Science, University of Waterloo\\
    Waterloo, ON N2L 3G1, Canada\\
    \email{aseem.baranwal@uwaterloo.ca}\\
    \email{shallit@uwaterloo.ca}
}

\maketitle

\begin{abstract}
In a recent paper of Rampersad et al., the authors conjectured that the smallest possible critical exponent of an infinite balanced word over a 5-letter alphabet is $3/2$.
We prove this result, using a formulation of first-order logic, the Pell number system, and a machine computation based on finite-state automata.
\end{abstract}

\section{Introduction}

In this paper, we prove a result about the critical exponent of infinite balanced words, using a formulation of first-order logic, the Pell number system, and a machine computation based on finite-state automata. To our knowledge, this is the first result in combinatorics on words to be proved using this approach via the Pell number system.

\subsection{Preliminaries}

Let $w$ denote a word over the alphabet $\Sigma$. If $w$ is finite, then $|w|$ denotes its length, and $|w|_a$ denotes the number of occurrences of the symbol $a$ in $w$, where $a \in \Sigma$. We let $\Fac(w)$ denote the set of all factors of $w$.

\begin{definition}
A word $w$ over the alphabet $\Sigma$ is {\em balanced} if for every symbol $a \in \Sigma$, and every pair of words $u$, $v \in \Fac(w)$ with $|u| = |v|$, we have $||u|_a - |v|_a| \le 1$.
\end{definition}

The class of \textit{Sturmian words} and the class of infinite aperiodic balanced words coincide over a binary alphabet. Vuillon~\cite{v2003} provides a survey on some previous work on balanced words, and Berstel et al.~\cite{bs2002} provide a survey on Sturmian words.

\begin{definition}
Let $w = w_0w_1 \cdots w_{n-1}$ be a finite word of length $n$. Then $p \in \mathbb{N}$ is a {\em period} of $w$ if $w_i = w_{i+p}$ for all $i$ with $0 \le i < n-p$.
\end{definition}

We say that a word $u$ has \textit{exponent} $e$ and write $u = z^e$, where $e = |u|/p$ is a positive rational number, and $z$ is the prefix of $u$ of length $p$; here $z$ is sometimes called the \textit{fractional root} of $u$. A word may have multiple periods, exponents, and fractional roots. We say $u$ is \textit{primitive} if its only integer exponent is $1$.  If $u$ is a finite nonempty word, then $u^\omega$ denotes the infinite word $uuu\cdots $.

\begin{example}
The word $w = \texttt{alfalfa}$ has three periods: $p_1=3$, $p_2=6$, and $p_3=7$. The corresponding exponents are $e_1=7/3$, $e_2=7/6$, and $e_3=1$. In this example, $w$ is a primitive word since its only integer exponent is $1$.
\end{example}

\begin{definition}
The {\em critical exponent} of an infinite word $\bf w$ is defined to be the supremum of the set of all rational numbers $r$ such that there exists a finite nonempty factor of $\bf w$ with exponent $r$. More formally,
$$E({\bf w}) = \sup\{r \in \mathbb{Q} : \text{there exist words } x,y \in \Fac({\bf w}) \text{ with } |y| > 0 \text{ and } y = x^r\}.$$
\end{definition}

\subsection{Previous work}
Rampersad et al.~\cite{rsv2018} gave a method to construct infinite balanced words from binary Sturmian words, using a characterization of recurrent aperiodic balanced words given by Hubert~\cite{h2000}. Their method is based on the notion of the constant gap property.

\begin{definition}
An infinite word ${\bf w}$ has the {\em constant gap property} if, for each symbol $a$, there is a positive integer $d$ such that the distance between successive occurrences of $a$ in ${\bf w}$ is always $d$.
\end{definition}

For example, $(0102)^\omega = 010201020102\cdots$ has the constant
gap property because the distance between consecutive $0$'s is
always 2, while the distance between consecutive $1$'s (reps., $2$'s)
is always 4.

Sturmian words ${\bf c}_{\alpha,\beta}$ can be defined in terms of two real parameters
$\alpha, \beta$ with $0 \leq \alpha, \beta < 1$, and $\alpha$ irrational.  Then
$${\bf c}_{\alpha,\beta}[n] := \lfloor \alpha (n+1) + \beta \rfloor - 
\lfloor \alpha n + \beta \rfloor.$$
A Sturmian word is called \textit{characteristic} if $\beta = 0$, and
is written as ${\bf c}_\alpha$.
In this case, it is well-known that an alternative characterization for these words can be given in terms of the continued fraction expansion of $\alpha = [d_0,d_1,d_2,\ldots]$ where $d_i \in \mathbb{N}$ for $i \geq 0$ and $d_i \geq 1$ for $i \ge 1$. 
Then ${\bf c}_\alpha$ is produced as the limit of the sequence of {\em standard words} $s_n$ defined as follows:
$$s_0 = 0, \quad s_1 = 0^{d_1-1}1, \quad  s_n = s_{n-1}^{d_n}s_{n-2} \text{ for } n \ge 2.$$

\begin{theorem}\label{thm1}
\cite{h2000} A recurrent aperiodic infinite word $\bf x$ is balanced if and only if $\bf x$ is obtained from a Sturmian word $\bf u$ over $\{0, 1\}$ by the following procedure: replace the $0$'s in $\bf u$ by a periodic sequence $\bf y$ with constant gaps over some alphabet $A$ and replace the $1$'s in $\bf u$ by a periodic sequence ${\bf y}'$ with constant gaps over some alphabet $B$, disjoint from $A$.
\end{theorem}

The authors of \cite{rsv2018} defined certain infinite balanced words ${\bf x}_k$ for $3 \le k \le 10$ constructed from a Sturmian characteristic word ${\bf c}_\alpha$, where $\alpha$, $\bf y$ and ${\bf y}'$ are carefully chosen. Table \ref{tab1} shows the choices for ${\bf x}_3$, ${\bf x}_4$, and ${\bf x}_5$. Here $\varphi = (1 + \sqrt{5})/2$ is the golden ratio. The authors also proved that $E({\bf x}_3) = 2 + \frac{\sqrt{2}}{2}$ and $E({\bf x}_4) = 1 + \frac{\varphi}{2}$;
furthermore, they showed that $E({\bf x}_3)$ is the least possible critical exponent over an alphabet of $3$ symbols. Based on computations, they also suggested that the least possible
critical exponents for balanced words over a $k$-letter
alphabet is $(k-2)/(k-3)$ for $k \ge 5$.   In this paper we take the first step towards this conjecture by proving the result for $k = 5$.
\begin{table}
\centering
\bgroup
\def\arraystretch{1.5}
\setlength\tabcolsep{5pt}
\begin{tabular}{|c|c|c|c|}
\hline
$k$ &  $\alpha$ & $y$ & $y'$\\
\hline
$3$ & $\sqrt{2}-1$ & $(01)^\omega$ & $2^\omega$\\
$4$ & $1/\varphi^2$ & $(01)^\omega$ & $(23)^\omega$\\
$5$ & $\sqrt{2}-1$ & $(0102)^\omega$ & $(34)^\omega$\\
\hline
\end{tabular}
\egroup
\vskip .1in
\caption{$\alpha$, $y$, and $y'$ used for the construction of ${\bf x}_k$.}\label{tab1}
\end{table}

\subsection{Automatic theorem proving using \texttt{Walnut}}
The authors in~\cite{rsv2018} employed a computational approach  using the automatic theorem-proving software \texttt{Walnut}~\cite{m2016}. The approach is based on the methods of Du et al.~\cite{dmss2016,mss2016}, using Theorems~\ref{thm2} and \ref{thm3} below. The $n^{\rm th}$ term of an arbitrary Sturmian characteristic word ${\bf c}_\alpha$, and consequently the generated infinite balanced word ${\bf x}_k$, can be computed by a finite automaton that takes the Ostrowski $\alpha$-representation~\cite{o1922} of $n$ as input.

\begin{theorem}\label{thm2} \cite[Theorem 9.1.15]{as2003}
Let $N \ge 1$ be an integer with Ostrowski $\alpha$-representation $b_j b_{j-1}\cdots b_0$. Then ${\bf c}_\alpha[N] = 1$ if and only if $b_j b_{j-1}\cdots b_0$ ends with an odd number of $0$'s.
\end{theorem}

\begin{theorem}\label{thm3} \cite[Theorem 12]{rsv2018}
Let $\alpha$ be a quadratic irrational and let ${\bf c}_\alpha$ be the Sturmian characteristic word with slope $\alpha$. Let $\bf x$ be any word obtained by replacing the $0$'s in ${\bf c}_\alpha$ with a periodic sequence $\bf y$ and replacing the $1$'s with a periodic sequence ${\bf y}'$. Then $\bf x$ is Ostrowski $\alpha$-automatic.
\end{theorem}

Using \texttt{Walnut}, we can constructively decide first-order predicates. When a predicate consisting of free variables is provided to \texttt{Walnut}, it also generates an automaton accepting values for the free variables that will satisfy the predicate. For predicates without any free variables, \texttt{Walnut} produces any of the two special automata, the \texttt{true}, and the \texttt{false} automaton, depending on whether the predicate is a tautology, or a contradiction respectively.

\section{Building the automata}
\label{automata}
We determine the critical exponent of ${\bf x}_5$ using the computational approach described above. The Ostrowski $\alpha$-numeration system for ${\bf x}_5$ is defined by the \textit{Pell} numbers, similar to how the numeration system for ${\bf x}_4$ is defined by the \textit{Fibonacci} numbers. To enable \texttt{Walnut} to work with this new numeration system, we require a deterministic finite automaton that reads its input in the Pell number system and recognizes the addition relation $\{(x, y, z) \in \mathbb{N}^3 : x+y=z\}$. Hieronymi and Terry~\cite{ht2018} showed that this is indeed possible when $\alpha$ is a quadratic irrational, and we have $\alpha = \sqrt{2}-1$ for ${\bf x}_5$, which satisfies this condition. Once we have the adder, a second automaton with output that can compute ${\bf x}_5$ is required for enabling \texttt{Walnut} to understand first-order predicates involving ${\bf x}_5$.

\subsection{Pell number system}
The Pell numbers are defined by the recurrence relation $P_n = 2P_{n-1} + P_{n-2}$ for $n \ge 2$ with $P_0 = 0$ and $P_1 = 1$.
The first few terms of this sequence are
$$ 0,1, 2, 5, 12, 29, 70, 169, \ldots $$
and form sequence \seqnum{A000129} in Sloane's {\it Encyclopedia} \cite{Sloane:2018}.  

We use this sequence of numbers to define a non-standard positional numeral system in the family of Ostrowski numeration systems~\cite{o1922} with $\alpha = \sqrt{2}-1$. Given an integer
$N$, we can express it as an integer linear combination of Pell
numbers as follows:   $N = \sum_{0 \leq i < n} d_i P_{i+1}$.
To ensure that this representation is unique, we impose the following conditions on the $d_i$:
\begin{enumerate}
    \item The least significant digit $d_0 \in \{0, 1\}$.
    \item For all $i>0$ we have $d_i \in \{0, 1, 2\}$.
    \item If $d_i=2$, then $d_{i-1} = 0$.
\end{enumerate}
In this case, the word $d_{n-1}d_{n-2}\cdots d_0$ is said to be the {\em canonical Pell representation} of an integer $N$, and we write it as $(N)_P$.

For example, $157$ has canonical Pell representation $(201100)_P$. Other representations of $157$ include $122100$, $201021$, and $122021$, but they do not conform to the conditions given above, and hence they are not canonical.

\subsection{Automaton for the addition relation in Pell-base}
To build the automaton that can recognize the addition relation in the Pell number system, we use Theorem \ref{thm4}, a corollary to the Myhill-Nerode theorem~\cite{n1958} based on the idea of Brzozowski derivative~\cite{b1964}.

\begin{definition}
Given a function $f: \Sigma^* \to \mathbb{R}$ for an alphabet $\Sigma$, we define its {\em Hankel matrix} $\mathcal{H} \in \mathbb{R}^{\Sigma^* \times \Sigma^*}$ as follows:
$$
\mathcal{H}
=
\bordermatrix{
    & \epsilon & a & b & aa & \cdots \cr
    \epsilon & f(\epsilon) & f(a) & f(b) & f(aa) & \cdots \cr
    a & f(a) & f(aa) & f(ab) & f(aaa) & \cdots \cr
    b & f(b) & f(ba) & f(bb) & f(baa) & \cdots \cr
    aa & f(aa) & f(aaa) & f(aab) & f(aaaa) & \cdots \cr
    \vdots & \vdots & \vdots & \vdots & \vdots & \ddots \cr
},
$$
where $\epsilon$ denotes the empty string. The matrix is indexed by words $u,v \in \Sigma^*$ such that $\mathcal{H}_{uv} = f(uv)$.
\end{definition}

\begin{theorem}\label{thm4}(Myhill-Nerode~\cite{n1958})
Let $\mathcal{L} = \{w_1, w_2, \ldots\}$ be a language over the finite alphabet $\Sigma$. Let $\mathcal{H}$ be a binary Hankel matrix indexed by the words $u, v \in \Sigma^*$ such that
$$ H_{uv} =
\begin{cases}
    1,  & \text{ if } uv \in \mathcal{L}; \\
    0,  & \text{ otherwise}.
\end{cases}
$$
Then $\mathcal{L}$ is regular if and only if the number of distinct rows in $\mathcal{H}$ is finite. Furthermore, the number of distinct rows equals the minimal number of states of a deterministic finite automaton recognizing $\mathcal{L}$.
\end{theorem}

For the indices of $\mathcal{H}$, we use a list of words over the alphabet $\Sigma_3 = \{0, 1, 2\}$ sorted in the \textit{radix} order. The radix order for two words $x$ and $y$ is defined by $x < y$, if $|x|<|y|$, or there exist symbols $a,b \in \Sigma$ such that $|x|=|y|$, $x=uax'$, $y=uby'$, and  $a<b$. The adder automaton takes as input $3$ integers $x, y, z$ in canonical Pell representation in parallel, and reaches an accepting state if and only if $x+y=z$. To achieve this, we require a generalization of the notion of Pell representation to $r$-tuples of integers for $r \ge 1$. A representation for $(x_1, x_2, \ldots, x_r)$ consists of a string of symbols $z$ over some alphabet $\Sigma$, such that a well-defined projection $\pi_i(z)$ over the $i^{\rm th}$ coordinate gives a canonical Pell representation of $x_i$. To handle this, first we pad the canonical Pell representations of smaller integers with leading $0$'s so that in the $r$-tuple, strings representing all $x_i$ have the same length. Our goal is to represent a triplet $(x, y, z)$ that will serve as an input to the adder automaton.

\begin{example}
Let $(x, y, z)=(65, 15, 80)$ be an integer triplet to be input to our adder automaton. We have $(65)_{10}=(020110)_P$, $(15)_{10}=(001011)_P$, and $(80)_{10}=(100200)_P$ after padding with sufficient leading $0$'s. Next, we project these representations as a series of ternary triplets, i.e., each digit belonging to the alphabet $\Sigma_3$. Hence $(65, 15, 80)$ is represented as
$$[0,0,1][2,0,0][0,1,0][1,0,2][1,1,0][0,1,0],$$
where the first digits of the triplets spell out $020110$, the canonical Pell representation of $65$. Similar claims about the second and third digits hold for $15$ and $80$, respectively. This projection is necessary to our method, because in order to recognize the addition relation, the automaton must be able to read all three integers $x, y, z$ in parallel.
\end{example}

Since each triplet has digits $\in \Sigma_3=\{0, 1, 2\}$, it follows that our input alphabet $\mathcal{P}$ for the adder automaton has size $3^3 = 27$. Next, we use a set of radix-ordered strings over $\mathcal{P}$ as indices for our binary Hankel matrix $\mathcal{H_P}$. The value in row $u$ and column $v$ of $\mathcal{H_P}$ ($u, v \in \mathcal{P}$) is $1$ if $uv$ denotes a series of triplets over $\Sigma_3$ that is a projection for an integer triplet $(x, y, z)$ such that $x+y=z$, and $0$ otherwise. Finally, we \textit{learn} the deterministic finite automaton with a combination of membership and equivalence queries using the \textit{Angluin} $L*$ algorithm~\cite{a1988}. The adder automaton contains $16$ states over an alphabet of size $27$, and hence it is infeasible to show here. The full automaton is publicly available on \href{https://github.com/aseemrb/Walnut/blob/master/Custom Bases/Visual/pell_adder.png}{GitHub}.\footnote{Corresponding \texttt{Walnut} code is available at \url{https://github.com/aseemrb/walnut}.}

Before we proceed, following an idea suggested to us by Luke Schaeffer \cite{dmss2016}, we prove the correctness of our adder automaton using \texttt{Walnut}. The proof is inductive, using the definition of successor of an integer. Of course the successor of an integer $x$ is $x+1$, but this makes use of the addition relation, which we have not yet proved. Instead we define the successor in a different way.
\begin{definition}
Given two integers $x$ and $y$, we say $y$ is the successor of $x$ if $x < y$, and  $(z \le x \text{ or } z \ge y)$ for all $z \in \mathbb{Z}$.
\end{definition}

The canonical Pell representation ensures that \texttt{Walnut} can perform comparisons on two integers in Pell representation easily, based on only their lexicographic ordering. Below we present the \texttt{Walnut} commands to compute the inductive proof. First, we define the \textit{successor} relation.

\begin{Verbatim}[numbers=left,xleftmargin=5mm,tabsize=2]
def pell_successor "?msd_pell
    x < y & (Az (z <= x) | (z >= y))";
\end{Verbatim}

The above command produces an automaton accepting pairs $(x, y)$ such that $y$ is a successor of $x$. \texttt{Walnut} also stores this definition and allows us to use it in other predicates. Next, we check the base case of the induction. For all $x, z \in \mathbb{Z}$, $x + 0 = z$ if and only if $x = z$. For the addition relation, $0$ is the identity element.

\begin{Verbatim}[numbers=left,xleftmargin=5mm,tabsize=2]
eval base_proof "?msd_pell Ax,z ((x + 0 = z) <=> (x = z))";
\end{Verbatim}

The predicate \texttt{base\_proof} produces the \texttt{true} automaton signifying that the predicate is true. We now verify our adder using the definition of successor. For all $x$, $y$, $z$, $u$, $v \in \mathbb{Z}$, if $u$ is the successor of $y$ and $v$ is the successor of $z$, then we have $x+y=z$ if and only if $x+u=v$.

\begin{Verbatim}[numbers=left,xleftmargin=5mm,tabsize=2]
eval inductive_proof "?msd_pell Ax,y,z,u,v
    ($pell_successor(y, u) & $pell_successor(z, v)) =>
    ((x + y = z) <=> (x + u = v))";
\end{Verbatim}

The predicate \texttt{inductive\_proof} also produces the \texttt{true} automaton. This completes the proof of correctness for our automaton recognizing the addition relation in Pell-base.   

\subsection{Automaton for computing ${\bf x}_5$}\label{automaton:x5}
\begin{figure}[htbp]
\centering
\includegraphics[width=0.4\textwidth]{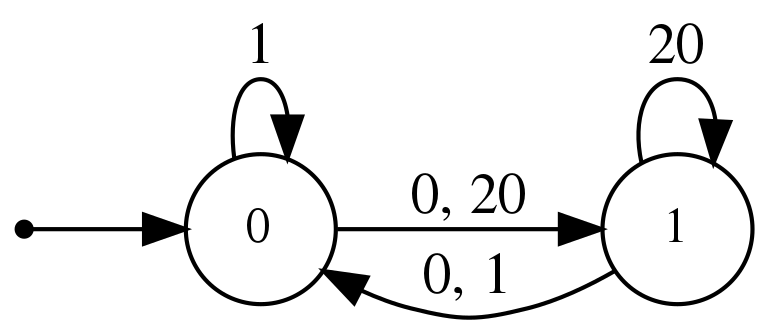}
\caption{Pell-base automaton for ${\bf c}_\alpha$.}
\label{c_alpha}
\end{figure}

Using the adder automaton we created above, we now build a deterministic finite automaton \textit{with output} that can compute ${\bf x}_5$. Each state of this automaton is associated with an output symbol from the alphabet $\Sigma_5 = \{0,1,2,3,4\}$ of ${\bf x}_5$. It takes as input an integer $N$ in canonical Pell representation, and halts at the state with output ${\bf x}_5[N]$. By Theorem \ref{thm3} we know that ${\bf x}_5$ is an $\alpha$-automatic sequence for $\alpha=\sqrt{2}-1$. Hence, using Theorem \ref{thm2} we first create an automaton for ${\bf c}_\alpha$, which is given by the limit of the sequence of finite words $s_n$ defined as follows:
$$s_0 = 0, \; s_1 = 01, \; s_n = s_{n-1}^2s_{n-2} \text{ for } n \ge 2.$$
This definition comes from the fact that the continued fraction expansion of $\alpha = \sqrt{2} -1$ is $[0,\bar{2}]$. The automaton given in Figure~\ref{c_alpha} produces the sequence ${\bf c}_\alpha$. The label on each state denotes the output associated with that state. When given a positive integer $N > 0$ as input, this automaton halts at the state with label ${\bf c}_\alpha[N]$. Here ${\bf c}_\alpha$ is indexed from $1$. The first few characters of ${\bf c}_\alpha$, and consequently of ${\bf x}_5$ constructed from ${\bf c}_\alpha$ are given below.

\begin{center}
{${\bf c}_\alpha = \texttt{01010010100101010010100101010}\cdots$ \\
${\bf x}_5 = \texttt{03140230410324031042301403240}\cdots$}
\end{center}

\begin{figure}[htbp]
\centering
\includegraphics[width=\textwidth]{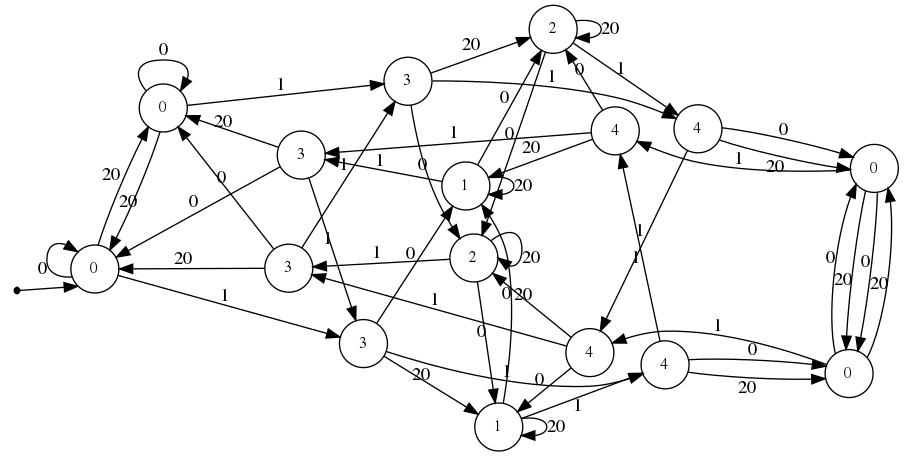}
\caption{Pell-base automaton for ${\bf x}_5$.}
\label{x5}
\end{figure}

Generating ${\bf x}_5$ from ${\bf c}_\alpha$ is a simple replacement of $0$'s and $1$'s by the constant-gap sequences ${\bf y} =(0102)^\omega$ and ${\bf y}'=(34)^\omega$ from Table \ref{tab1}. We start indexing ${\bf x}_5$ from $0$. Let ${\bf z}_l$ denote the prefix of ${\bf c}_\alpha$ with length $l$. Then for $(i \ge 0)$, the value of ${\bf x}_5[i]$ is is a function of ${\bf c}_\alpha[i+1]$, $|{\bf z}_{i+1}|_0$, and $|{\bf z}_{i+1}|_1$. Figure~\ref{x5} shows the Pell-base automaton that generates the word ${\bf x}_5$. The labels on the states denote the output symbol for that state. Recall that the canonical Pell representation for an integer cannot end in a $2$, and cannot have a $2$ immediately followed by a $1$ or $2$ based on the restrictions we have imposed. For this reason, we display certain transitions in the automaton to consist of two symbols. This is done to remove those intermediate states from display that do not produce any output. In practice, the automaton will never halt at such states because the input is always a valid canonical Pell representation.

As an example, consider the canonical Pell representation of $(25)_{10}$, which is $(2001)_P$. When the automaton is given the input string \texttt{2001}, it halts at a state with label $3$, signifying that ${\bf x}_5[25]=3$.

We now use \texttt{Walnut} to verify that the sequence produced by the automaton in Figure~\ref{x5} conforms to the definition of ${\bf x}_5$ given by Rampersad et al.~\cite{rsv2018}. The proof is based on the definition of constant gap words ${\bf y} = (0102)^\omega$ and ${\bf y}' = (34)^\omega$. Recall that ${\bf c}_\alpha$ is indexed from $1$, while ${\bf x}_5$ is indexed from $0$. In subsequent \texttt{Walnut} commands, let \texttt{C} denote the automaton given in Figure~\ref{c_alpha}, and \texttt{X} denote the automaton in Figure~\ref{x5}. We start by verifying the replacement of $0$'s. The first symbol of ${\bf c}_\alpha$ is $0$, and it is replaced by a $0$ in ${\bf x}_5$.
\begin{Verbatim}[numbers=left,xleftmargin=5mm,tabsize=2]
eval first_0_to_0 "?msd_pell C[1] = @0 & X[0] = @0";
\end{Verbatim}
The second occurrence of 0 in ${\bf c}_\alpha$ at position 3, is replaced by a 1 in ${\bf x}_5$.
\begin{Verbatim}[numbers=left,xleftmargin=5mm,tabsize=2]
eval second_0_to_1 "?msd_pell C[3] = @0 & X[2] = @1";
\end{Verbatim}
For any three $0$'s occurring in ${\bf c}_\alpha$ at positions $p, q, r$ such that $p < q < r$, and all other positions between $p$ and $r$ are occupied by $1$'s, the replacement must be one of the sequences: $010$, $102$, $020$, or $201$.
\begin{Verbatim}[numbers=left,xleftmargin=5mm,tabsize=2]
eval possible_triplets_for_0s "?msd_pell Ap,q,r
    ((p < q) & (q < r) &
     (C[p + 1] = @0) &
     (C[q + 1] = @0) &
     (C[r + 1] = @0) &
     (Ai ((i > p) & (i < r) & (i != q)) =>
         (C[i + 1] = @1))) =>
    (((X[p] = @0) & (X[q] = @1) & (X[r] = @0)) |
     ((X[p] = @1) & (X[q] = @0) & (X[r] = @2)) |
     ((X[p] = @0) & (X[q] = @2) & (X[r] = @0)) |
     ((X[p] = @2) & (X[q] = @0) & (X[r] = @1)))";
\end{Verbatim}
Next, we verify the replacement of $1$'s. The first occurrence of $1$ in ${\bf c}_\alpha$ at position $2$ is replaced by a $3$ in ${\bf x}_5$.
\begin{Verbatim}[numbers=left,xleftmargin=5mm,tabsize=2]
eval first_1_to_3 "?msd_pell C[2] = @1 & X[1] = @3";
\end{Verbatim}
Any two $1$'s in ${\bf c}_\alpha$ that have all $0$'s in between must be replaced by alternate $3$ and $4$.
\begin{Verbatim}[numbers=left,xleftmargin=5mm,tabsize=2]
eval alternate_3_4_for_1s "?msd_pell Ap,q
    ((p < q) &
     (C[p + 1] = @1) &
     (C[q + 1] = @1) &
     (Ai ((i > p) & (i < q)) => (C[i + 1] = @0))) =>
    (((X[p] = @3) & (X[q] = @4)) |
     ((X[p] = @4) & (X[q] = @3)))";
\end{Verbatim}
All the predicates above produce the \texttt{true} automaton. This completes the verification of the automaton for ${\bf x}_5$.

\section{Writing the proof}
The authors in~\cite{rsv2018} proposed a hypothesis about the critical exponent of ${\bf x}_5$, but were not able to prove it. We prove their hypothesis using the automata created in Section~\ref{automata}. The hypothesis is that the critical exponent of the infinite balanced word ${\bf x}_5$ is $E({\bf x}_5)=3/2$. The predicates used to prove this hypothesis do not contain free variables. In \texttt{Walnut}, such predicates evaluate to either the \texttt{true} or the \texttt{false} automaton. Please see~\cite{m2016} for further details.

\subsection{Proving the hypothesis}
We complete the proof in three steps. First, we test whether there exist integers $i, n, p$ such that a length-$n$ factor of ${\bf x}_5$ starting at index $i$ and having period $p$ has exponent at most $3/2$. This predicate produces the \textit{true} automaton.

\begin{Verbatim}[numbers=left,xleftmargin=5mm,tabsize=2]
eval fac_low_exponent "?msd_pell Ei,p,n
    (p >= 1) & (2*n <= 3*p) & (Aj (j + p < n) =>
    X[i + j] = X[i + j + p])";
\end{Verbatim}

Next, we test whether there exist integers $i, n, p$ such that a length-$n$ factor of ${\bf x}_5$ starting at index $i$ and having period $p$ has exponent exactly equal to $3/2$. This predicate also produces the \textit{true} automaton.

\begin{Verbatim}[numbers=left,xleftmargin=5mm,tabsize=2]
eval fac_ex_exponent "?msd_pell Ei,p,n
    (p >= 1) & (2*n = 3*p) & (Aj (j + p < n) =>
    X[i + j] = X[i + j + p])";
\end{Verbatim}

Finally, we test whether there exist integers $i, n, p$ such that a length-$n$ factor of ${\bf x}_5$ starting at index $i$ and having period $p$ has exponent greater than $3/2$. This predicate \texttt{fac\_high\_exponent} produces the \textit{false} automaton.

\begin{Verbatim}[numbers=left,xleftmargin=5mm,tabsize=2]
eval fac_high_exponent "?msd_pell Ei,p,n
    (p >= 1) & (2*n > 3*p) & (Aj (j + p < n) =>
    X[i + j] = X[i + j + p])";
\end{Verbatim}

Combining the results above, we conclude that there exists a factor of ${\bf x}_5$ with exponent $=3/2$ and there does not exist a factor of ${\bf x}_5$ with exponent $>3/2$. Hence, the critical exponent of ${\bf x}_5$ is $3/2$.

\subsection{Exploring interesting properties}
Although the proof is complete, we can go a step further and explore interesting properties of the word ${\bf x}_5$ using our method. For example, in order to find the factors of ${\bf x}_5$ that have exponent exactly $=3/2$, we use the following command.

\begin{Verbatim}[numbers=left,xleftmargin=5mm,tabsize=2]
eval fac_cex5 "?msd_pell En
    (p >= 1) & (2*n = 3*p) & (Aj (j + p < n) =>
    X[i + j] = X[i + j + p])";
\end{Verbatim}

\begin{figure}[htbp]
\centering
\includegraphics[width=\textwidth]{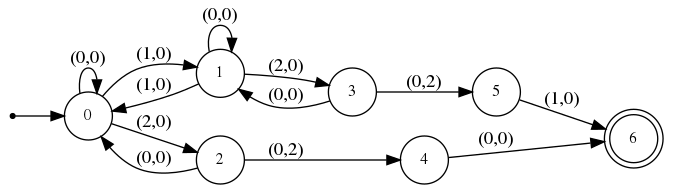}
\caption{Pairs $(i, p)$ such that factors of ${\bf x}_5$ with starting index $i$ and period $p$ have exponent $=3/2$.}
\label{exact}
\end{figure}

Note that this predicate has two free variables $i$ and $p$. The corresponding automaton is given in Figure~\ref{exact}, which accepts pairs of integers $(i, p)$, such that there exists a factor $w$ of ${\bf x}_5$ with period $p$, starting at index $i$. The automaton suggests that for all such pairs $(i, p)$, we have $p=4$. We state this observation as Corollary~\ref{cex5:length}, and present Example~\ref{ex3} as an illustration.

\begin{corollary}\label{cex5:length}
All factors $w$ of ${\bf x}_5$ with exponent $E({\bf x}_5) = 3/2$ have length $|w|=6$ and period $p=4$.
\end{corollary}

\begin{example}\label{ex3}
Consider the pair $(23, 4)$ whose projection is $[1,0][2,0][0,0][1,0]$. This sequence is accepted by the automaton in Figure~\ref{exact}. We look at the factors of ${\bf x}_5$ starting at position $23$. The factor which is the maximal power starting at this position is $w =\texttt{403240}$, which indeed has the exponent $3/2$.
\end{example}

Another interesting property to explore could be the possible periods $p$, for which a factor of ${\bf x}_5$ is ``almost" a $3/2$-power. There are many ways to define this property. To formalize this, let $w = zz'$ be a factor of ${\bf x}_5$ with length $n$, where $z'$ is a prefix of $z$. The period of $w$ is $p = |z|$. For $p > 10$, we define a factor to be an ``almost" $3/2$-power if $n \ge 3p/2 - 2$.

\begin{Verbatim}[numbers=left,xleftmargin=5mm,tabsize=2]
eval almost_ce_period "?msd_pell Ei
    (p > 10) &
    (2*n + 4 >= 3*p) &
    (Aj (j + p < n) => X[i + j] = X[i + j + p])";
\end{Verbatim}

\begin{figure}[htbp]
\centering
\includegraphics[width=4.5in]{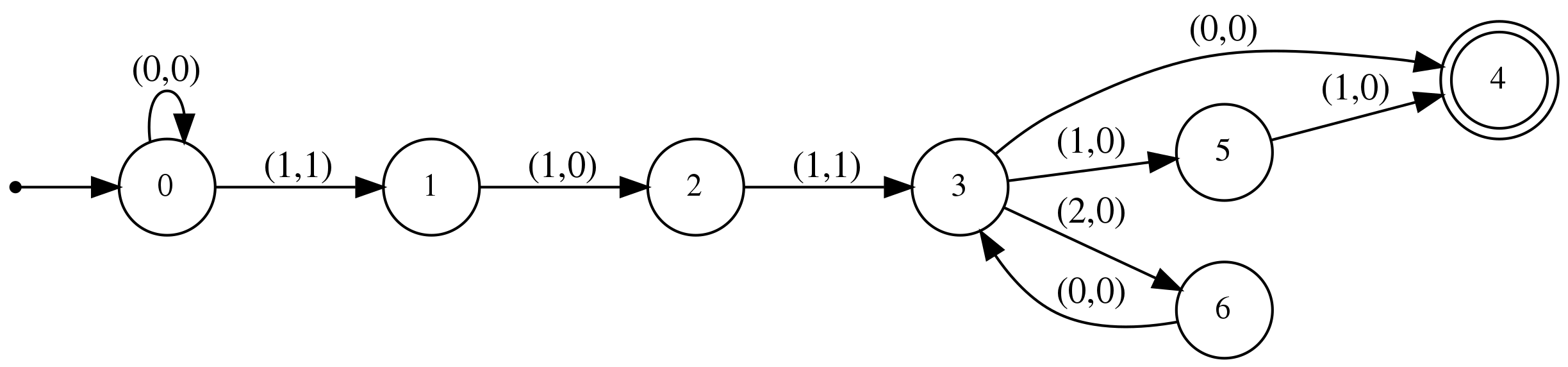}
\caption{Pairs $(n, p)$ characterizing factors of ${\bf x}_5$ that are ``almost" $3/2$-powers.}
\label{almost}
\end{figure}

Note that this predicate has free variables $n$ and $p$, which indicate the length and period of $w$ respectively. The automaton produced for this predicate is shown in Figure~\ref{almost}. We observe that for $p > 10$, all pairs $(n, p)$ have the form:

$$\binom{1}{1} \binom{1}{0} \binom{1}{1} \bigg \{ \binom{2}{0} \binom{0}{0} \bigg \}^* \binom{0}{0}, \text{ or}$$
$$\binom{1}{1} \binom{1}{0} \binom{1}{1} \bigg \{ \binom{2}{0} \binom{0}{0} \bigg \}^* \binom{1}{0} \binom{1}{0}.$$

This shows that there exist infinitely many factors of ${\bf x}_5$ with this property. We also note that, as $p$ approaches infinity, the exponent of these factors approaches $3/2$, which is the critical exponent of ${\bf x}_5$.

\section{Breadth-first search}
We have proved the existence of
a balanced word over $\Sigma_5 = \{ 0,1,2,3,4 \}$ of critical exponent $3/2$.

It now remains to show that this exponent $3/2$ is optimal for the alphabet $\Sigma_5$.  To do this, we use a computer program that employs the usual breadth-first search technique.  We use the following simple observations to narrow the search space:  first, we assume the first letter is $0$.  Second, we impose the restriction that the first occurrence of the letter $i$ occurs before the first occurrence of $j$ if $i < j$.
With these restrictions, the longest balanced word of critical exponent $< 3/2$ is of length $44$, and there are exactly $5$ of them:
\begin{center}
{\tt 01203104120130410213014021031401203104120130 \\
01203240210320421023042012302401203240210320\\
01230240120324021032042102304201230240120324\\
01231421023124102132412013214201231421023124\\
01231430132143103213410312341301231430132143}
\end{center}

\section{Future prospects}
\subsection{Other words characterized by Pell-base}
\begin{figure}[htbp]
\centering
\includegraphics[width=3.5in]{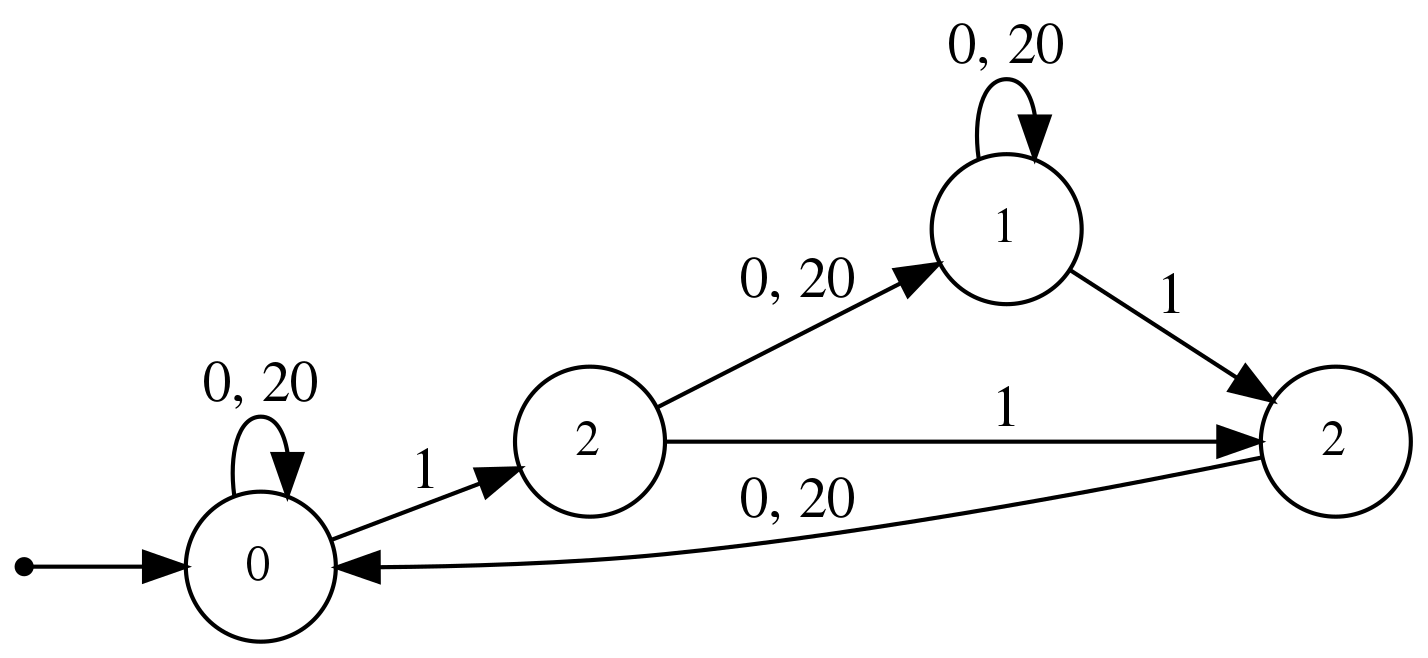}
\caption{Pell-base automaton for ${\bf x}_3$.}
\label{x3}
\end{figure}

The authors in~\cite{rsv2018} determine the value of the critical exponent of the infinite balanced word ${\bf x}_3$, $E({\bf x}_3) = 2 + \sqrt{2}/2$ using a manual case-based proof. Given that the value of $\alpha=\sqrt{2}-1$ for ${\bf x}_3$ is the same as that for ${\bf x}_5$, we can easily determine the value of $E({\bf x}_3)$ using our method. The constant gap words used for constructing ${\bf x}_3$ from ${\bf c}_\alpha$ are ${\bf y}=(01)^\omega$ and ${\bf y}'=2^\omega$ (see Table~\ref{tab1}). Using the same procedure as described in Section~\ref{automaton:x5}, we build an automaton with output that produces the sequence ${\bf x}_3$ (see Figure~\ref{x3}). The claimed value in this case is irrational, unlike $E({\bf x}_5) = 3/2$, which means that it is never actually attained by any factor of ${\bf x}_3$.

In the subsequent \texttt{Walnut} commands, let \texttt{X} denote the automaton in Figure~\ref{x3}. First, we compute the periods $p$ such that a repetition with exponent $\ge 13/5$ and period $p$ occurs in ${\bf x}_3$.
\begin{Verbatim}[numbers=left,xleftmargin=5mm,tabsize=2]
eval periods_of_high_powers "?msd_pell Ei
    (p >= 1) & (Aj (5*j <= 8*p) => X[i + j] = X[i + j + p])";
\end{Verbatim}
The language accepted by the produced automaton is $0^*110000^*$, which is the Pell-base representation of numbers of the form $P_n + P_{n-1}$, for $n \ge 5$. The following command lets us save this as a regular expression.
\begin{Verbatim}[numbers=left,xleftmargin=5mm,tabsize=2]
reg pows msd_pell "0*110000*";
\end{Verbatim}
Next, we compute pairs of integers $(n, p)$ such that ${\bf x}_3$ has a factor of length $n+p$ with period $p$, and this factor cannot be extended to a longer factor of length $n+p+1$ with the same period.
\begin{Verbatim}[numbers=left,xleftmargin=5mm,tabsize=2]
def maximal_reps "?msd_pell Ei
    (Aj (j < n) => X[i + j] = X[i + j + p]) &
    (X[i + n] != X[i + n + p])";
\end{Verbatim}
Finally, we compute integer pairs $(n, p)$ where $p$ is of the form $0^*110000^*$, and $n+p$ is the maximum possible length of any factor with period $p$.
\begin{Verbatim}[numbers=left,xleftmargin=5mm,tabsize=2]
eval highest_powers "?msd_pell
    (p >= 1) & $pows(p) & $maximal_reps(n, p) &
    (Am $maximal_reps(m, p) => m <= n)";
\end{Verbatim}
The predicate \texttt{highest\_powers} produces an automaton accepting integer pairs $(n , p)$ that have the form
$$\binom{0}{0}^* \binom{2}{1} \binom{0}{1} \binom{2}{0} \binom{0}{0} \bigg \{ \binom{2}{0} \binom{0}{0} \bigg \}^* \binom{0}{0}, \text{ or}$$
$$\binom{0}{0}^* \binom{2}{1} \binom{0}{1} \binom{2}{0} \binom{0}{0} \bigg \{ \binom{2}{0} \binom{0}{0} \bigg \}^* \binom{1}{0} \binom{1}{0}.$$
As clear from the pattern of pairs $(n, p)$, for $m \ge 5$, when $p=P_m + P_{m-1}$, then we have $n = P_{m+1} - 2$. Thus, we have the exponent as the ratio of length $n+p$ and period $p$,
\begin{equation}\label{eqn:exp}
e=\frac{P_{m+1}+P_m+P_{m-1}-2}{P_m+P_{m-1}} = 2 + \frac{P_m-2}{P_m+P_{m-1}}.
\end{equation}
Let $a_k/b_k = [d_0,d_1,d_2,\ldots,d_k]$ be the \textit{convergents} of $\alpha=\sqrt{2}-1$. Then for $k \ge 0$, we have $a_k=P_k$ and $b_k=P_{k+1}$. Hence, the following bound holds:
\begin{equation}\label{eqn:conv}
\bigg|\alpha-\frac{P_k}{P_{k+1}}\bigg| < \frac{1}{P_{k+1}P_{k+2}} < \frac{1}{P_{k+1}^2}.
\end{equation}
Substituting $k=m-1$ in~(\ref{eqn:conv}), we bound the value of $P_m/P_{m-1}$. We also know that $P_m/P_{m-1}$ converges to the the \textit{silver ratio}, $\sigma=\sqrt{2}+1$. Thus, we have,
$$e=2+\frac{P_m-2}{P_m+P_{m-1}}<2+\frac{\sqrt{2}+1+1/P_{m-1}^2-2/P_{m-1}}{\sqrt{2}+2-1/P_{m-1}^2}.$$
For $m \ge 5$, as $m\rightarrow\infty$, the value of $e$ is increasing, and tends to $2+\sqrt{2}/2$. Thus, $e<2+\sqrt{2}/2$, which proves the hypothesis.

In theory, one might also hope to determine the values of $E({\bf x}_8)$ and $E({\bf x}_9)$ using our method, since the corresponding continued fraction expansions of $\alpha$ end in a repeating $2$. The actual result depends on the practical limitations of run-time and memory availability for the corresponding machine computation.

\subsection{Open problems}
The obvious open problem to pursue is implementation of an automaton that recognizes the addition relation for words in the general Ostrowski $\alpha$-numeration system (see~\cite{ht2018}). Assuming that the machine computation is feasible, we might be able to obtain analogous proofs for balanced words over larger alphabets.

\section{Acknowledgments}

We thank Narad Rampersad and Luke Schaeffer for their helpful comments.

%
%
\bibliographystyle{splncs04}
\bibliography{abbrevs,references}

\end{document}